\documentclass[prl,twocolumn,showpacs]{revtex4}
\usepackage[dvips]{graphicx}
\usepackage{latexsym}
\usepackage{bm}
\begin{document}
\newcommand{\fig}[2]{\includegraphics[width=#1]{#2}}
\newcommand{\pprl}{Phys. Rev. Lett. \ }
\newcommand{\pprb}{Phys. Rev. {B}}
\newcommand{\be}{\begin{equation}}
\newcommand{\ee}{\end{equation}}
\newcommand{\bea}{\begin{eqnarray}}
\newcommand{\eea}{\end{eqnarray}}
\newcommand{\nn}{\nonumber}
\newcommand{\la}{\langle}
\newcommand{\ra}{\rangle}
\newcommand{\dg}{\dagger}
\newcommand{\upa}{\uparrow}
\newcommand{\dna}{\downarrow}
\newcommand{\nco}{Na$_x$CoO$_2$ \ }

\title{Charge and spin order on the triangular lattice --- Na$_x$CoO$_2$
at $x=0.5$}

\author{Sen Zhou and Ziqiang Wang}
\affiliation{Department of Physics, Boston College, Chestnut Hill,
MA 02467}

\date{\today}

\begin{abstract}

The nature of electronic states due to strong correlation and
geometric frustration on the triangular lattice is investigated
in connection to the unconventional insulating state of Na$_x$CoO$_2$
at $x=0.5$. We study an extended Hubbard model
using a spatially unrestricted Gutzwiller
approximation. We find a new class of charge and spin ordered
states at $x=1/3$ and $x=0.5$ where antiferromagnetic (AF) frustration
is alleviated via weak charge inhomogeneity.
At $x=0.5$, we show that the $\sqrt{3}a\times2a$ off-plane Na dopant order
induces weak $\sqrt{3}a\times1a$ charge order in the Co layer. The symmetry
breaking enables successive $\sqrt{3}a\times1a$ AF and $2a\times2a$
charge/spin ordering transitions at low temperatures. The Fermi surface
is truncated by the $2a\times2a$
hexagonal zone boundary into small electron and hole pockets.
We study the phase structure and compare to recent experiments.
%
%\typeout{polish abstract}
\end{abstract}
\maketitle

Sodium doped cobaltate Na$_x$CoO$_2$ has emerged recently as an important,
layered triangular lattice fermion system with
a rich phase structure \cite{foo}.
These include a $5$K superconducting phase near $x=1/3$ upon
hydration \cite{takada03}; an A-type antiferromagnetic (AF) phase around
$x=0.8$ \cite{keimer05}; and an unexpected insulating state at
$x=0.5$ \cite{foo}. A series of experiments find the insulating state
unconventional. While the magnetic susceptibility
shows two cusps at $T_{m1}=88$K and $T_{m2}=53$K, the in-plane
resistivity exhibits only a derivative feature at $T_{m1}$, followed
by a metal-insulator transition below $T_{m2}$ \cite{foo}. The insulating
state has a small optical gap of $15$meV \cite{nlwang,hwang} and an
anisotropic single-particle gap of $\sim8$meV in angle resolved
photoemission spectroscopy (ARPES) \cite{hasangap}.
Electron diffraction \cite{zandbergen,huang}
shows that Na orders into $\sqrt{3}a\times2a$ (hereafter we set
$a=1$) supercells below $\sim300$K,
suggesting that dopant order induced charge order may
play a role in the insulating behaviors \cite{foo,nlwang,kwlee}. However,
NMR experiments show that the Co valence exhibits small disproportionation
with no appreciable change across the metal-insulator
transition \cite{bobroff}. Bobroff et al. proposed that
the insulating state is a result of successive SDW
transitions due to the crossing of the Fermi surface (FS) with the orthorhombic
zone boundary of $\sqrt{3}\times2$ Na order \cite{bobroff}.
Recently, elastic neutron scattering discovered that
AF order occurs at $T_{m1}$ with a $2\times2$ hexagonal unit cell
\cite{younglee}. The ordering vector is clearly incompatible with
and challenges the SDW scenario.

In this paper, we study theoretically the electronic state at
$x=0.5$. The relevant low energy
electronic structure involves three Co $t_{2g}$ atomic orbitals forming
one $a_{1g}$ and two $e_g^\prime$ bands in the solid.
In a recent work \cite{zhouetal}, starting from the three-band
Hubbard model of the $t_{2g}$ complex with LDA band dispersions
\cite{singh00}, it is shown that strong correlation renormalizes
the crystal field splitting and the bandwidths and
drives the $e_g^\prime$ band below $E_F$, leaving
a single band of mostly $a_{1g}$ character near the Fermi level.
The resulting quasiparticle dispersion and FS topology are in
agreement with ARPES over a wide range of Na doping
\cite{hbyang2,hasan2,hbyang1} as well as in hydrated samples
\cite{arpes-hydrated}. This justifies a single-band model for the
basic low energy physics, provided that the strong
Coulomb repulsion is included at the Co site.

We consider here a single-band $t$-$U$-$V$ model and study the interplay
between the frustration of the
kinetic energy and the AF spin correlations. Specifically, we extend the
Gutzwiller approximation to the variational space spanned by spatially
unrestricted and spin dependent densities. We find that the
tendency towards inhomogeneity due to strong correlation and magnetic
frustration work together to alleviate the AF frustration and produce
a class of charge and spin ordered states. Hereafter, we use the
terms inhomogeneity, charge and spin order interchangeably to refer
to a nonuniform electronic state where the densities of charge and spin
are spatially and periodically modulated. At $x=1/3$, the ground state
has spontaneous $\sqrt{3}\times\sqrt{3}$ charge and spin order even
when $V=0$. The frustration is avoided as the AF order develops
on the honeycomb lattice and coexists with weak
charge density modulations. At $x=0.5$, we find that a large $V$ is necessary
to destabilize the uniform paramagnetic phase towards a state
with $\sqrt{3}\times 1$ charge and AF spin order.
This state is close to a Wigner crystal with a large charge disproportionation
and a large insulating gap \cite{ogata,choyetal}, inconsistent with
NMR \cite{bobroff}, transport \cite{foo,nlwang,hwang},
and ARPES \cite{hasangap} experiments.
We show that the $\sqrt{3}\times2$ Na order at $x=0.5$
induces a weak $\sqrt{3}\times1$ charge order at high temperatures.
The symmetry breaking allows $\sqrt{3}\times1$ AF order to develop
at $T_{m1}$. Remarkably, the FS at $x=0.5$ coincides well with
the $2\times2$ hexagonal zone boundary. This allows the $2\times2$
charge and spin order to develop by umklapp scattering
at a lower temperature $T_{m2}$. The truncation of the FS into
small electron and hole pockets marks the onset of the insulating behavior.

We begin with the one-band $t$-$U$-$V$ model
\be
H=\sum_{ij\sigma}t_{ij}c_{i\sigma}^\dagger c_{j\sigma}
+U\sum_i \hat n_{i\uparrow}\hat n_{i\downarrow}+ V\sum_{i>j}
{\hat n_i \hat n_j\over |\vec r_i-\vec r_j\vert},
\label{h}
\ee
where, $c_{i\sigma}^\dagger$ creates an $a_{1g}$ {\it hole}
of spin $\sigma$, $\hat n_i$ is
the hole density operator, $U$ and $V$ are the on-site and long-range Coulomb
repulsion. From the LDA $a_{1g}$ band dispersion, the hopping parameters
are chosen according to $t_{ij}=(-202,35,29)$meV for the first,
second, and third nearest neighbors respectively.
The local electron doping density is given by $x_i=1-n_i$.
The large-$U$ limit of Eq.~(\ref{h}) is usually treated in the Gutzwiller
approximation (GA) \cite{vollhard,fczhang}, which corresponds to the saddle
point of the slave-boson path integral formulation \cite{kr}.
Since the superexchange interaction is very small in the
cobaltates due to the small bandwidth and large-U,
we neglect the AF Heisenberg interaction \cite{keimer05} and
consider magnetism of a kinetic origin.
To encompass the Hilbert space with inhomogeneous
charge/spin densities, we adopt a spatially unrestricted GA described
by the renormalized mean-field Hamiltonian
\bea
H_{GA}&=&\sum_{ij\sigma} g_{ij}^\sigma t_{ij} c_{i\sigma}^\dagger c_{j\sigma}
+\sum_{i,\sigma} \varepsilon_{i\sigma} (c_{i\sigma}^\dagger c_{i\sigma}
-n_{i\sigma})
\nonumber \\
&+&V\sum_{i>j}
{\hat n_i \hat n_j\over |\vec r_i-\vec r_j\vert},
\label{hgw}
\eea
where the Gutzwiller renormalization factor $g_{ij}^\sigma$ depends
on the sites connected by the hopping integral $t_{ij}$,
% and on the spin of the hopping electron,
\be
g_{ij}^\sigma=\sqrt{x_i x_j\over(1-n_{i\sigma})(1-n_{j\sigma})}.
\label{g}
\ee
The $\varepsilon_{i\sigma}$ in Eq.~(\ref{hgw}) is a spin dependent
local fugacity that maintains the equilibrium condition and
local densities upon Gutzwiller projection
\cite{lietal,qhwang}. It is determined by
$\partial \langle H_{GA} \rangle/\partial n_{i\sigma}=0$.
The spin density is $S_i^z=(n_{i\uparrow}-n_{i\downarrow})/2$.
If the charge density is forced to be uniform,
the solution of Eq.~(\ref{hgw}) gives a stable paramagnetic phase
up to $x_0\simeq0.67$ where a ferromagnetic (FM) transition
takes place. The AF order, typical of large-U systems at small $x$
on bipartite lattices \cite{kr}, is absent due to geometrical frustration.
We show below that this frustration
can be alleviated by forming inhomogeneous electronic states.
\begin{figure}
\begin{center}
\fig{3.0in}{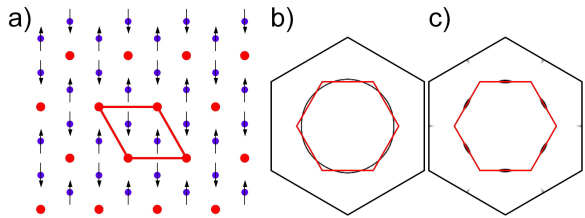}
\vskip-2.4mm
\caption{
(a) $\sqrt{3}\times\sqrt{3}$ charge and spin order at $x=1/3$ and $V=0$.
The charge and spin densities are $x=(0.32,0.36,0.32)$ and
$S^z=(0.18,0.00,-0.18)$. (b) Nesting of
$\sqrt{3}\times\sqrt{3}$ zone boundary with the paramagnetic FS.
(c) Intensity of the quasiparticle peaks at the Fermi level.
}
\label{fig1}
\end{center}
\vskip -9mm
\end{figure}

To this end, we
consider large triangular lattices of $240\times320$ sites with $6\times8$
unit cells wherein $x_i$, $n_{i\sigma}$, and $\varepsilon_{i\sigma}$ are
allowed to have spatial variations. They are determined
self-consistently by standard iterations. First let's consider the case
at $x=1/3$. It is remarkable that even for $V=0$ the ground
state has spontaneous $\sqrt{3}\times\sqrt{3}$
charge and spin order displayed in Fig.~1a.
Frustration is alleviated as AF moments
reside on the underlying unfrustrated honeycomb lattice.
The tendency to avoid AF frustration is
materialized because the FS at $x=1/3$ coincides with the
$\sqrt{3}\times\sqrt{3}$ zone boundary, as shown in Fig.~1b,
such that {\it weak} charge order develops by the ``umklapp'' scattering
and anisotropic gapping of the FS shown in Fig.~1c.
We find that charge and spin sectors are coupled in the sense that
the removal of inhomogeneity in either will reinstate the uniform paramagnetic
phase. This state is thus different from the $\sqrt{3}\times\sqrt{3}$
charge ordered state proposed by Motrunich and Lee \cite{motrunich,ogata13}
near $x=1/3$, which is a Wigner crystal due to Coulomb jamming
under a large $V$, involving no spin ordering and a large insulating
gap. In contrast, our charge and spin ordered state has
a very small insulating gap at $V=0$ which increases gradually
with increasing $V$. We point out that such a spin/charge ordered
state has not yet been observed at $x=1/3$, most likely because
of the Na dopant disorder \cite{singh-na}.
Indeed, we find that a disordered Na potential of moderate strength
destroys the long-range order.
It will be interesting to examine whether enhanced AF fluctuations
in proximity to the ordered state can lead to superconductivity.

Next we turn to $x=0.5$, which is not a natural commensurate
filling on the triangular lattice. As a result, the uniform paramagnetic
state is found to be stable for small $V<V_c$, $V_c\simeq1.35$eV.
For $V>V_c$, we find a first order transition
to an inhomogeneous state with charge/spin order. Fig.~2 shows that, for
$V=1.5$eV, the moments are ordered into unfrustrated AF chains that
are AF coupled and separated by nonmagnetic chains.
This state is driven by strong long-range Coulomb
$V$ and stabilized by AF spin correlations. The charge density has
a $\sqrt{3}\times1$ unit cell with strong
disproportionation close to the Co$^{3+}$/Co$^{4+}$ configuration,
resulting in a large charge gap. Indeed, similar ``AF Wigner
crystal'' was recently proposed \cite{choyetal} and
independently investigated by
variational Monte Carlo on frustrated lattices \cite{ogata}.
It is important to note that while the
$\sqrt{3}\times1$ AF order can be described by a unit cell
that is equivalent to the $2\times2$ hexagonal unit cell
deduced from neutron scattering, the structure factor is different and
the magnetic Bragg peaks shown in Fig.~2 (green squares) only has
two-fold symmetry. This differs from
the hexagonal Bragg peaks observed in neutron scattering,
unless 120 degree orientated domains of equal contribution are present
\cite{younglee}. For this reason, we will continue to refer to
this spin pattern as $\sqrt{3}\times1$ AF order (see also Fig.~3b)
and reserve the term $2\times2$ hexagonal unit cell for a
state with hexagonal Bragg peaks (see Fig.~3c).
In view of the experimental findings of weak charge
disproportionation by NMR \cite{bobroff}, and small insulating gaps
by transport, optics, and ARPES \cite{foo,nlwang,hasangap},
we conclude that this state, though indicative of the structure of
the AF order by avoiding frustration via inhomogeneity, cannot describe the
$x=0.5$ phase of the cobaltates.
\begin{figure}
\begin{center}
\fig{2.8in}{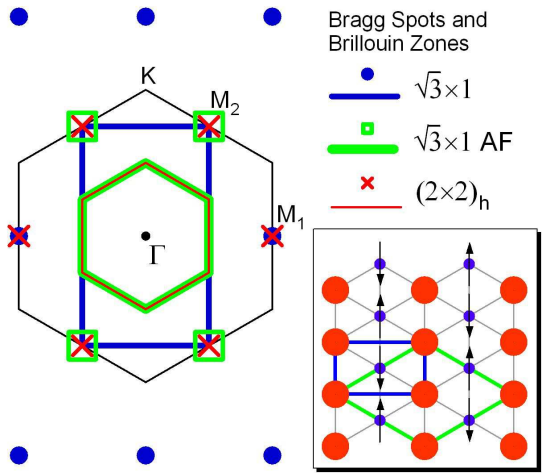}
\vskip-2.4mm
\caption{
$\sqrt{3}\times1$ charge and AF spin order at $x=0.5$ and
$V=1.5$eV (lower right). The charge and spin densities:
$x=(0.05,0.95,0.05)$ and $S^z=(0.48,0.00,-0.48)$.
Also shown are charge and magnetic
zones and Bragg peak locations of different structures.
Note the absence of Bragg spot at $M_1$ for $\sqrt{3}\times1$ AF order.
$(2\times2)_h$ corresponds to the hexagonal magnetic zone and Bragg
spots of the state shown in Fig.~3c.
}
\label{fig2}
\end{center}
\vskip-9mm
\end{figure}

It turns out that the Na dopant order plays an important but subtle role
at $x=0.5$. Below about $300$K, Na orders into $\sqrt{3}\times2$
superlattice structures \cite{zandbergen,huang}.
This has led to the notion of an induced $\sqrt{3}\times2$
electron charge order in the Co plane,
which is ultimately responsible for the insulating state at $x=0.5$.
Three remarks are in order. First, the electrostatic potential
from off-plane ordered ionic dopants
in transition metal oxides is usually not strong due to screening by phonons,
interband transitions, and the mobile carriers. Thus, the induced
charge order in the basal plane is at most moderate.
Second, as the temperature is lowered, FS stability
may occur depending on the charge ordering symmetry. The latter can
be different from the symmetry of dopant order.
Due to the fact that the zigzag chains of
ordered Na dopants above and below a Co layer are staggered, the
superlattice potential felt by the electrons has a higher symmetry
and results in $\sqrt{3}\times1$ charge order with a much elongated
orthorhombic zone than originally thought.
Third, the breaking of the lattice symmetry makes it energetically favorable
to develop AF order by alleviating frustration via weak charge inhomogeneity.
\begin{figure}
\begin{center}
\fig{3.3in}{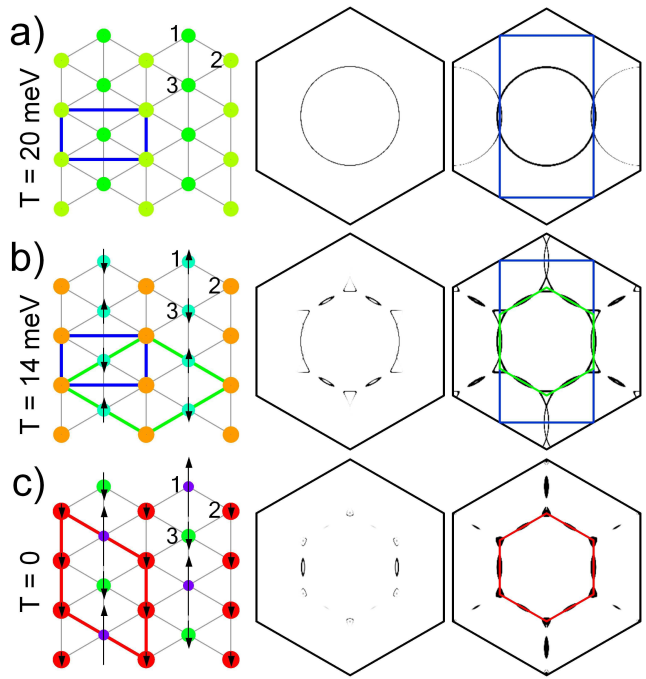}
\vskip-2.4mm
\caption{
The self-consistent states at $x=0.5$ for three different temperatures
(a), (b), and (c). First column: charge/spin ordering
patterns and the unit cells. Second column: FS without thermal
broadening showing the anisotropic gapping of the FS.
Third column: FS when intensity scale is reduced by four orders of
magnitude showing the band folding along zone boundaries
of corresponding charge/spin order.
}
\label{fig3}
\end{center}
\vskip-9mm
\end{figure}

The effect of the Na dopant potential is to add
\be
V_{\rm dopant}(i)=V_d\sum_{I=1}^{N_{\rm Na}}{\hat n_i\over\sqrt
{\vert\vec r_I-\vec r_i\vert^2
+d_z^2}}
\label{vdopant}
\ee
to Eq.~(\ref{hgw}), where $V_d$ is the potential strength
and $d_z \simeq a$ is the setback distance of Na to the Co plane.
Including carrier screening of the dopant potential,
we set $V=0.2$eV and $V_d=0.5$eV in the calculation.
The Gutzwiller renormalized mean-field theory in Eq.~(\ref{hgw})
enables the calculation of the free-energy at finite temperatures.
We present the temperature evolution of the self-consistently
determined states in Fig.~3, and that of the charge and spin density
in Fig.~4. At a temperature $k_BT=20$meV above $T_{m1}$, marked in
Fig.~4, the electronic state shown in Fig.~3a has $\sqrt{3}\times1$
charge order induced by the Na potential {\it without} magnetic order.
Note that the charge modulation shown in Fig.~4 is
weak ($\sim2\%$) and the FS in Fig.~3a is not
affected. Lowering the intensity scale by four orders of magnitude reveals
the weak band folding patterns along the intersections of the FS with
the $\sqrt{3}\times1$ zone boundary.
As the temperature is reduced, we find an AF transition at
$T_{m1}$ marked in Fig.~4, below which
$\sqrt{3}\times1$ AF order develops and coexists with
the $\sqrt{3}\times1$ charge order. This state is shown
in Fig.~3b at $k_BT=14$meV. The transition is primarily a spin ordering
transition, involving small changes in the charge disproportionation
($\sim3\%$ from Fig.~4). This is consistent with
NMR measurements \cite{bobroff}. Remarkably, Fig.~3b shows that
the FS at $x=0.5$ coincides well with the hexagonal magnetic zone boundary.
However, since the structure factor of the $\sqrt{3}\times1$ AF order
breaks the hexagonal symmetry, the vertical sections of
the FS remain intact as the scattering between them
is switched off by the absence of the magnetic Bragg peak
at $M_1$ in Fig.~2. Thermal broadening
and the 120 degree domains may further weaken the features of $T_{m1}$ in
resistivity, optics, and ARPES measurements. Nevertheless, the
FS topology change due to the emergence of electron-like FS pockets
leads to the reduction and sign change in the Hall
coefficient observed below $T_{m1}$ \cite{foo}.

\begin{figure}
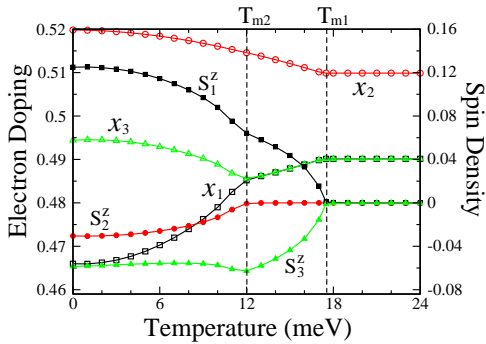

\begin{center}
\vskip+2mm
\fig{2.5in}{fig4.eps}
\vskip-1.4mm
\caption{
Temperature evolution of charge (electron doping) and
spin densities at three sites marked in Fig.~3. The spin/charge
ordering transitions are marked by $T_{m1}$ and $T_{m2}$.
}
\label{fig4}
\end{center}
\vskip-9mm
\end{figure}

Reducing the temperature further, we find a second
transition marked as $T_{m2}$ in Fig.~4 associated with additional symmetry
breaking. Below $T_{m2}$, small magnetic moments develop at the Co(2) sites.
They are coupled {\it ferromagnetically} as shown in Fig.~3c at $T=0$.
Below $T_{m2}$, the charge and spin densities in Fig.~4
are all different on the three inequivalent Co sites.
Hence the ground state has charge and spin order with identical
$2\times2$ hexagonal unit cell. Moreover, the Fourier transform
of the spin density has now hexagonal Bragg peaks
in a single domain (see Fig.~2) of similar weight at low temperatures.
As a result of the combined $2\times2$ charge/spin order, the umklapp
scattering destroys almost the entire FS as shown in Fig.~3c,
leading to the onset of the metal-insulator transition below $T_{m2}$.
The residual electron and hole-like FS pockets have
six-fold symmetry with a two-fold anisotropy and provide
an explanation for the existence of small FS pockets observed
by Shubnikov-de Haas oscillations \cite{balicas05}. The low temperature
insulating behavior develops by the localization of the electronic
states on the small FS pockets, such that the insulating state has a
nonzero density of states, and in this sense, a pseudogap.
We emphasize that the most important character of the transition
at $T_{m2}$ is the emergence of small FM moments at Co(2).
Its ordering direction should not be taken literally since it is
specific to the restriction to collinear spins in the Gutzwiller approach.
The FM moment on Co(2), collinear with the AF moments on Co(1)
and Co(3), causes magnetic frustration and leads to an imbalance of the
AF moments shown in Fig.~3c. In fact,
it is more energetically favorable for the small FM moment to point
orthogonal to the AF moment so as to avoid frustration of the in-plane
AF order. The physics of the second transition is, however, unchanged.
The small FM moments have been observed by NMR experiments \cite{imai,yokoi}
as associated with the $53$K transition, but they may be too small to have
been detected in the neutron scattering experiments \cite{younglee}.

To summarize, we have shown that the unconventional insulating state
at $x=0.5$ is a result of the interplay among strong correlation,
$\sqrt{3}\times1$ weak charge order induced by Na
dopant order, AF order by alleviated frustration,
and the overlap of the FS with the $2\times2$ hexagonal
magnetic zone boundary.
A single band $t$-$U$-$V$ model including the Na dopant potential
for the electron doped, hole-like Co $a_{1g}$ band on the triangular
lattice captures the basic physics of the charge and spin order.
The transition temperatures $T_{m1}$ and $T_{m2}$, the
size of the moments, and the insulating gap will
depend on microscopic details
%such as the band parameters
%and the strength of the dopant potential $V_d$, and thus must
and vary when Na is replaced by other isoelectronic atoms such as potassium
in K$_{0.5}$CoO$_2$. However, the phase structure discussed here is expected to
be universal of the cobaltate family at $x=0.5$.

We thank H. Ding, Y. S. Lee, and especially P.A. Lee for many valuable
discussions. This work is supported by DOE grant DE-FG02-99ER45747 and
ACS grant 39498-AC5M.

\end{document}